\newcommand{\be}{\begin{eqnarray}}
\newcommand{\ee}{\end{eqnarray}}
\documentstyle[aps,epsfig]{revtex}

\begin{document}
\draft
\title{Pion dispersion relation at finite temperature in the linear sigma 
       model from chiral Ward identities}
\author{Alejandro Ayala, Sarira Sahu}
\address{Instituto de Ciencias Nucleares\\
         Universidad Nacional Aut\'onoma de M\'exico\\
         Aptartado Postal 70-543, M\'exico Distrito Federal 04510, M\'exico.}
\author{Mauro Napsuciale}
\address{Instituto de F\'{\i}sica\\
         Universidad de Guanajuato\\
         Loma del Bosque 103, Le\'on Guanajuato 37160, M\'exico.}
\maketitle
\begin{abstract}

We develop one-loop effective vertices and propagators in the linear sigma
model at finite temperature satisfying the chiral Ward identities. We use 
these in turn to compute the pion dispersion relation in a pion medium for 
small momentum and temperatures on the order of the pion mass at next to 
leading order in the parameter $m_\pi^2/4\pi^2f_\pi^2$ and to zeroth order in 
the parameter $(m_\pi/m_\sigma)^2$. We show that this expansion reproduces the
result obtained from chiral perturbation theory at leading order. The main 
effect is a perturbative, temperature-dependent increase of the pion mass. We 
find no qualitative changes in the pion dispersion curve with respect to the 
leading order behavior in this kinematical regime.

\end{abstract}

The propagation properties of pions in a thermal hadronic environment are a 
key ingredient for the proper understanding of several physical processes 
taking place in dense and hot plasmas. The scenarios include dense 
astrophysical objects such as neutron stars --where the inclusion of 
mesonic degrees of freedom is essential to determine the equation of state of 
matter in the inner core-- and the evolution of the highly interacting region 
formed in the aftermath of a high-energy heavy-ion collision. In the latter 
context, knowledge of the pion dispersion relation would help to properly 
account for the component of the dilepton spectrum originated in the hadronic 
phase, below the critical temperatures and densities for the formation of a 
quark-gluon plasma (QGP)~\cite{Gale}, and to shed light on possible collective 
surface phenomena which are speculated to occur as a consequence of the mainly
attractive interaction between pions and nuclei~\cite{Shuryak}.

In order to account for the hadronic degrees of freedom at temperatures
and densities below the QCD phase transition to a QGP, one needs to resort to 
effective chiral theories whose basic ingredient is the fact that pions are
Goldston bosons, originated in the spontaneous breakdown of chiral symmetry.
Chiral perturbation theory ($\chi$PT) is one of such effective theories that 
has been successfully employed to show the well known result that at leading 
perturbative order and at low momentum, the modification of the pion
dispersion curve in a pion medium is just a constant, temperature dependent,
increase of the pion mass~\cite{Gasser}. Any deviation from this
leading order behavior in perturbation theory can only be looked for at higher
orders. Second order calculations of the pion dispersion curve can be found in 
Refs.~\cite{other,Goity}.    

The simplest realization of chiral symmetry is nevertheless provided by the 
much studied linear sigma model which possesses the convenient feature of
being a renormalizable field theory, both at zero~\cite{Lee} and at finite 
temperature~\cite{Mohan}. This model has lately received a boost of
attention in view of recent theoretical results ~\cite{Tornqvist} and 
analyses of data~\cite{Svec} that seem to confirm, though not without
controversy~\cite{Isgur-Harada}, that a broad scalar resonance, with a mass
in the vicinity of 600 MeV --that can be identified with the $\sigma$-meson-- 
indeed exists.

In addition to the above mentioned characteristics, the linear sigma model 
also reproduces --as we will show-- the leading order modification to the pion
mass in a thermal pion medium (for temperatures on the order of the pion 
mass), when use is made of a systematic expansion in the ratio 
$(m_\pi /m_\sigma)^2$ at zeroth order, where $m_\sigma$, $m_\pi$
are the vacuum sigma and pion masses, respectively. This important point, 
often missed in the literature~\cite{Contreras}, provides the connection 
between the linear sigma model and $\chi$PT at finite temperature. 

In this work, we use the linear sigma model to construct effective vertices 
and propagators to one loop. We use these in turn to compute the one loop 
modification to the pion propagator in a pion medium. We require that the
effective vertices and propagators thus constructed satisfy the chiral Ward 
identities~\cite{Lucio}. The net result is a next to leading order calculation
in the parameter $m_\pi^2/4\pi^2f_\pi^2$, where $f_\pi$ is the pion decay 
constant and to zeroth order in the parameter $(m_\pi /m_\sigma)^2$. From the
effective pion propagator, we explore the behavior of the pion dispersion 
curve at small momentum and for temperatures on the order of the pion mass. 
Though the momentum dependence of the correction term is non-trivial, we find 
no qualitative difference from the leading order result in this regime.
Possible effects due to a high nucleonic density as well as general details of
the calculation are left out for a future work.

The Lagrangian for the linear sigma model, including only the mesonic degrees 
of freedom and after the explicit inclusion of the chiral symmetry breaking 
term, can be expressed as~\cite{Lee}
\be
   {\mathcal{L}}=\frac{1}{2}\left[(\partial{\mathbf{\pi}})^2 +
                (\partial\sigma)^2 - m_\pi^2{\mathbf{\pi}}^2 - 
                m_\sigma^2\sigma^2\right] - 
                \lambda^2 f_\pi\sigma (\sigma^2 + {\mathbf{\pi}}^2) -
                \frac{\lambda^2}{4}(\sigma^2 + {\mathbf{\pi}}^2)^2\, ,
   \label{lagrangian}
\ee
where $\mathbf{\pi}$ and $\sigma$ are the pion and sigma fields, respectively,
and the coupling $\lambda^2$ is given by
\be
   \lambda^2=\frac{m_\sigma^2-m_\pi^2}{2f_\pi^2}\, .
   \label{coupling}
\ee

From the Lagrangian in Eq.~(\ref{lagrangian}), one obtains the Green's
functions and the Feynman rules to be used in perturbative calculations, in the
usual manner. Thus, in particular, the bare pion and sigma propagators 
$\Delta_\pi (P)$, $\Delta_\sigma (Q)$ and the bare one-sigma 
two-pion and four-pion vertices $\Gamma_{12}^{ij}$, $\Gamma_{04}^{ijkl}$ are 
given by (hereafter, capital letters are used to denote four momenta)
\be
   i\Delta_\pi(P)\delta^{ij}&=&\frac{i}{P^2-m_\pi^2}\delta^{ij}\nonumber \\
   i\Delta_\sigma (Q)&=&\frac{i}{Q^2-m_\sigma^2}\nonumber \\
   i\Gamma_{12}^{ij}&=&-2i\lambda^2 f_\pi\delta^{ij}\nonumber \\
   i\Gamma_{04}^{ijkl}&=&-2i\lambda^2(\delta^{ij}\delta^{kl} + 
                          \delta^{ik}\delta^{jl} + \delta^{il}\delta^{jk})\, .
   \label{rules}
\ee
These Green's functions are sufficient to construct the modification to the 
pion propagator, both at zero and finite temperature, at any given 
perturbative order. 

An alternative approach consists on exploiting the relations 
that chiral symmetry imposes among different n-point Green's functions. These
relations, better known as chiral Ward identities ($\chi$WIs), are a direct
consequence of the fact that the divergence of the axial current may be used
as an interpolating field for the pion. For example, two of the $\chi$WIs
satisfied --order by order in perturbation theory-- by the functions 
$\Delta_\pi (P)$, $\Delta_\sigma (Q)$, $\Gamma_{12}^{ij}$ and 
$\Gamma_{04}^{ijkl}$ are~\cite{Lee}
\be
   f_\pi\Gamma_{04}^{ijkl}(;0,P_1,P_2,P_3)&=&
   \Gamma_{12}^{kl}(P_1;P_2,P_3)\delta^{ij} + 
   \Gamma_{12}^{lj}(P_2;P_3,P_1)\delta^{ik} +
   \Gamma_{12}^{jk}(P_3;P_1,P_2)\delta^{il}\nonumber \\
   f_\pi\Gamma_{12}^{ij}(Q;0,P)&=&
   \left[\Delta_\sigma^{-1}(Q) - 
   \Delta_\pi^{-1}(P)\right]\delta^{ij}\, ,
   \label{Ward}
\ee
where momentum conservation at the vertices is implied, that is
$P_1+P_2+P_3=0$ and $Q+P=0$. Therefore, any perturbative modification of one of
these functions introduces modifications in other, when the former are related 
to the latter through  $\chi$WIs. 

At one loop and after renormalization, the sigma propagator is modified by 
finite terms. At zero temperature this modification is purely imaginary and
its physical origin is that a sigma particle, with a mass larger than twice 
the mass of the pion, is unstable and has a (large) non-vanishing width coming
from its decay channel into two pions. At finite temperature the modification
results on real and imaginary parts. The real part modifies the sigma
dispersion curve whereas the imaginary part represents a temperature dependent
contribution to the sigma width. 
%%%%%%%%%%%%%%%%%%%%%%%%%%%%%%%%%%%%%%%%%%%%%%%%%%%%%%%%
\begin{figure}[h!] % fig 1
\hspace{-2.5cm}
\centerline{\epsfig{file=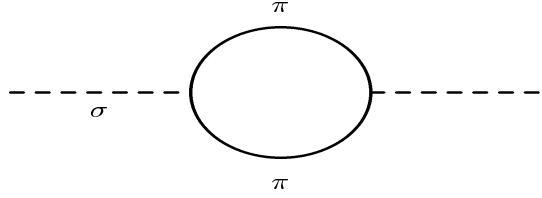,height=1.0in,width=1.5in}
}
\vspace{1cm}
\caption{
Dominant contribution to the sigma self-energy at one loop.
}
\end{figure}
%%%%%%%%%%%%%%%%%%%%%%%%%%%%%%%%%%%%%%%%%%%%%%%%%%%%%%%%

After renormalization, the leading one-loop contribution to the sigma 
self-energy, in an expansion in the parameter $(m_\pi/m_\sigma )^2$, can be 
shown~\cite{progress} to arise from the diagram depicted in Fig.~1. In the 
imaginary-time formalism of Thermal Field Theory (TFT), the expression for 
this diagram is given as $6\lambda^4f_\pi^2I(\omega,q)$, where the function 
$I$ is defined by
\be
   I(\omega,q)=T\sum_n\int\frac{d^3k}{(2\pi)^3}
   \frac{1}{\omega_n^2 + k^2 +m_\pi^2}\ 
   \frac{1}{(\omega_n - \omega )^2 + 
   ({\mathbf{k}} - {\mathbf{q}})^2 + m_\pi^2}\, .
   \label{sigmaself}
\ee 
Here $\omega = 2m\pi T$ and $\omega_n = 2n\pi T$ ($m$, $n$ integers) are
discrete boson frequencies and $q=|{\mathbf{q}}|$. From Eq.~(\ref{sigmaself})
we obtain the time-ordered version $I^t$ of the function $I$, after 
analytical continuation to Minkowski space. The imaginary part of $I^t$ is 
given by
\be
   {\mbox I}{\mbox m}I^t(q_0,q)&=&
   \frac{\varepsilon (q_o)}{2i}\left[
   I(i\omega\rightarrow q_o + i\epsilon ,q) - 
   I(i\omega\rightarrow q_o - i\epsilon ,q)\right]
   \nonumber \\
   &=&-\frac{1}{16\pi}
   \left\{ a(Q^2)+\frac{2T}{q}\ln\left(\frac{1-e^{-\omega_+(q_0,q) /T}}
   {1-e^{-\omega_-(q_0,q) /T}}\right)\right\}
   \Theta(Q^2-4m_\pi^2)\, ,
   \label{imtimeorder}
\ee
where $Q^2=q_0^2-q^2$, $\varepsilon$ and $\Theta$ are the sign and step 
functions, respectively, and the functions $a$ and $\omega_\pm$ are given by
\be
   a(Q^2)&=&\sqrt{1-\frac{4m_\pi^2}{Q^2}}\nonumber \\
   \omega_\pm (q_0,q)&=&\frac{|q_0| \pm a(Q^2)q}{2}\, ,
   \label{imfunc}
\ee
whereas the real part of $I^t$ at $Q=0$ is given by
\be
   {\mbox R}{\mbox e}I^t(0)=-\frac{1}{8\pi^2}\int_0^\infty
   \frac{dk}{E_k}\left[1+2f(E_k)\right]\, ,
   \label{refunc}
\ee
where $E_k=\sqrt{k^2+m_\pi^2}$ and the function $f$ is the Bose-Einstein 
distribution
\be
   f(E_k)=\frac{1}{e^{E_k/T}-1}\, .
   \label{be}
\ee
Therefore, the one-loop effective sigma propagator becomes
\be
   i\Delta_\sigma^\star (Q)&=&\frac{i}{Q^2 - m_\sigma^2 + 
                              6\lambda^4f_\pi^2I^t(Q) }\, .
   \label{newsigprop}
\ee
The temperature-independent infinities are absorbed into the redefinition of 
the physical masses and coupling constants by the introduction of suitable
counterterms in the usual manner. 

In order to preserve the $\chi$WIs expressed in Eq.~(\ref{Ward}), the 
corresponding one-loop effective one-sigma two-pion and four-pion vertices
become
\be
   i\Gamma_{12}^{\star\ ij}(Q;P_1,P_2)&=&-2i\lambda^2 f_\pi\delta^{ij}
   \left[1 - 3\lambda^2I^t(Q)\right]\nonumber \\
   i\Gamma_{04}^{\star\ ijkl}(;P_1,P_2,P_3,P_4)&=&-2i\lambda^2\left\{
   \left[1 - 3\lambda^2I^t(P_1+P_2)\right]\delta^{ij}\delta^{kl} \right.
   \nonumber\\
   &+& 
   \left[1 - 3\lambda^2I^t(P_1+P_3)\right]\delta^{ik}\delta^{jl} 
   \nonumber\\
   &+& \left.
   \left[1 - 3\lambda^2I^t(P_1+P_4)\right]\delta^{il}\delta^{jk}\right\}\, .
   \label{newvertices}
\ee
It can be shown~\cite{progress} that the functions in Eq.~(\ref{newvertices}) 
arise from considering all of the possible one-loop contributions to the 
one-sigma two-pion and four-pion vertices, when maintaining only the zeroth 
order terms in a systematic expansion in the parameter $(m_\pi/m_\sigma )^2$.
%%%%%%%%%%%%%%%%%%%%%%%%%%%%%%%%%%%%%%%%%
\vspace{1cm}
\begin{figure}[h!] % fig 2
\hspace{-1.5cm}
\centerline{\epsfig{file=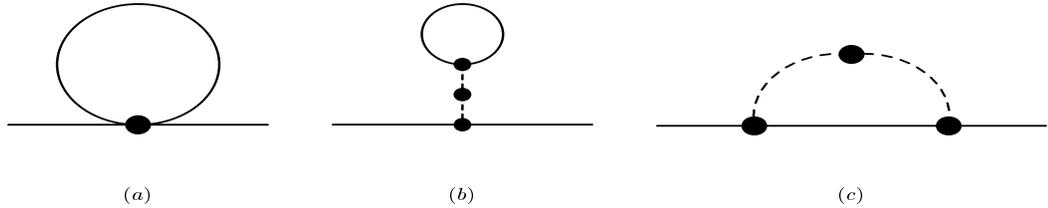,height=1.0in,width=4.5in}
}
\vspace{1cm}
\caption{
Dominant contributions to the pion self-energy at one loop in
the effective expansion. The heavy dots denote the effective vertices and
propagators.
}
\end{figure}
%%%%%%%%%%%%%%%%%%%%%%%%%%%%%%%%%%%%%%%%%

We now use the above effective vertices and propagator to construct the
one-loop modification to the pion self-energy. The leading contributions come
from the diagrams depicted in Fig.~2. These are, explicitly
\be
   \Pi_a(P)\ \delta^{ij}&=&\lambda^2\delta^{ij}\int_\beta\frac{d^4K}{(2\pi)^4}
   \frac{1}{K^2-m_\pi^2}\left\{5-9I^t(0)-6\lambda^2I^t(P+K)\right\}
   \nonumber\\
   \Pi_b(P)\ \delta^{ij}&=&-\lambda^2\delta^{ij}
   \left(\frac{6\lambda^2f_\pi^2}{m_\sigma^2}\right)\int_\beta
   \frac{d^4K}{(2\pi)^4}\frac{1}{K^2-m_\pi^2}
   \left\{\frac{[1-3\lambda^2I^t(0)]^2}{[1-\frac{6\lambda^4f_\pi^2}{m_\sigma^2}
   I^t(0)]}\right\}
   \nonumber\\
   \Pi_c(P)\ \delta^{ij}&=&-\lambda^2\delta^{ij}
   \left(\frac{4\lambda^2f_\pi^2}{m_\sigma^2}\right)\int_\beta
   \frac{d^4K}{(2\pi)^4}\frac{1}{K^2-m_\pi^2}
   \left\{\frac{[1-3\lambda^2I^t(P+K)]^2}
   {[1-\frac{6\lambda^4f_\pi^2}{m_\sigma^2}
   I^t(P+K)-\frac{(P+K)^2}{m_\sigma^2}]}\right\}\, ,
   \label{selfenergies}
\ee
where the subindex $\beta$ means that the integrals are to be computed at
finite temperature. We now expand the denominators in the second and third
of Eqs.~(\ref{selfenergies}), keeping only the leading order contribution
when considering $m_\pi$, $T$ and $P$ as small compared to $m_\sigma$. Adding
up the above three terms and to zeroth order in $(m_\pi/m_\sigma)^2$ where
\be
   \lambda^2\left(1-\frac{2\lambda^2f_\pi^2}{m_\sigma^2}\right)\approx
   \frac{m_\pi^2}{2f_\pi^2}\, , 
   \label{approx}
\ee
the pion self-energy can be written as
\be
   \Pi(P)&=&\left(\frac{m_\pi^2}{2f_\pi^2}\right)
   T\sum_n\int\frac{d^3k}{(2\pi)^3}\frac{1}{K^2+m_\pi^2}
   \left\{5 + 2\left(\frac{P^2+K^2}{m_\pi^2}\right)\right.\nonumber\\
   &-&\left.\left(\frac{m_\pi^2}{2f_\pi^2}\right)
   \left[9I^t(0) + 6I^t(P+K)\right]\right\}\, ,
   \label{self}
\ee
where we carry out the calculation in the imaginary-time formalism of TFT with
$K=(\omega_n,{\mathbf{k}})$ and $P=(\omega,{\mathbf{p}})$. The pion dispersion
relation is thus obtained from the solution to
\be
   P^2 + m_\pi^2 +{\mbox R}{\mbox e}\Pi(P)=0\, ,
   \label{dispersion}
\ee
after the analytical continuation $i\omega\rightarrow p_o + i\epsilon$. 

As it stands, Eq.~(\ref{self}) contains a temperature-dependent infinity 
coming from the product of the vacuum piece in ${\mbox R}{\mbox e}I^t(0)$ and 
the temperature-dependent piece in the indicated integral, as well as vacuum 
infinities. However, it is well known that finite temperature does not 
introduce new divergences and that whatever regularization and renormalization 
is needed at zero temperature, it will also be necessary and sufficient at 
finite temperature. Let us recall that at one loop, renormalization is carried 
out by introducing appropriate counterterms into the original Lagrangian and 
that at this level, no temperature-dependent infinities arise. Going to two 
loops, one will always encounter temperature-dependent infinities in integrals 
involving only the bare terms of the original Lagrangian. The remarkable
property of renormalizable theories is that these temperature-dependent 
infinities are exactly canceled by the contribution from the loops computed by 
using the counterterms that were introduced at one loop. The above was 
explicitly shown for the case of the self-energy in the $\phi^4$ theory by 
Kislinger and Morley~\cite{Kislinger} and for the sigma vacuum expectation 
value in the linear sigma model by Mohan~\cite{Mohan}. For this reason, here 
we omit the renormalization procedure and consider only the finite, 
temperature-dependent terms throughout the rest of the calculation and refer
the reader to the cited references for details.

Let us first look at the dispersion relation at leading order. After analytical
continuation all the terms are real. The integrals involved are 
\be
   T\sum_n\int\frac{d^3k}{(2\pi)^3}\frac{1}{K^2+m_\pi^2}&\rightarrow&
   \frac{1}{2\pi^2}\int_0^\infty\frac{dk\ k^2}{E_k}f(E_k)\nonumber\\
   &\equiv&\frac{m_\pi^2}{2\pi^2}\ g(T/m_\pi)\nonumber\\
   T\sum_n\int\frac{d^3k}{(2\pi)^3}\frac{K^2}{K^2+m_\pi^2}&\rightarrow&
   -\ m_\pi^2\ \left(\frac{m_\pi^2}{2\pi^2}\right)\ g(T/m_\pi)\, ,
   \label{g}
\ee 
where $g$ is a dimensionless function of the ratio $T/m_\pi$ and the arrows
indicate only the temperature dependence of the expressions. Thus, the 
dispersion relation results from
\be
   \left[1 + 2\ \xi\ g(T/m_\pi)\right](p_0^2-p^2) -
   \left[1 + 3\ \xi\ g(T/m_\pi)\right]m_\pi^2 = 0\, ,
   \label{dis1a}
\ee
with $\xi=m_\pi^2/4\pi^2f_\pi^2\ll 1$. For $T\sim m_\pi$, $g(T/m_\pi)\sim1$,  
therefore, at leading order and in the kinematical regime that we are 
considering, Eq.~(\ref{dis1a}) can be written as 
\be
   p_0^2 = p^2 + m_\pi^2\left[1 + \xi\ g(T/m_\pi)\right]\, ,
   \label{dis1b}
\ee
which coincides with the result obtained from $\chi$PT~\cite{Gasser}. 

We now look at the next to leading order terms in Eq.~(\ref{self}). The first
of these is purely real and represents a constant, second order shift to the
pion mass squared 
\be
   -\ 9\left(\frac{m_\pi^2}{2f_\pi^2}\right)^2
   T\sum_n\int\frac{d^3k}{(2\pi)^3}\frac{I^t(0)}{K^2+m_\pi^2}
   \rightarrow
   \frac{9}{2}\ \xi^2\
   g(T/m_\pi)\ h(T/m_\pi)\ m_\pi^2\, ,
   \label{secordmass}
\ee
where $h$ is a dimensionless function of the ratio $T/m_\pi$ defined by
\be
   h(T/m_\pi)\equiv\int_0^\infty\frac{dk}{E_k}f(E_k)\, .
   \label{h}
\ee
The remaining term in Eq.~(\ref{self}) shows a non-trivial dependence on $P$.
It involves the function $S$ defined by
\be
   S(P)\equiv T\sum_n\int\frac{d^3k}{(2\pi)^3}\frac{I^t(P+K)}{K^2+m_\pi^2}\, .
   \label{s}
\ee
The sum is performed by resorting to the spectral representation of $I^t$ and 
$(K^2+m_\pi^2)^{-1}$. Thus, the real part of the retarded version of $S$, 
after analytical continuation is
\be
   {\mbox R}{\mbox e}S^r(p_0,p)&\equiv&\frac{1}{2}
   \left[S(i\omega\rightarrow p_0+i\epsilon ,p) + 
   S(i\omega\rightarrow p_0-i\epsilon ,p)\right]\nonumber\\
   &=&\ -\ {\mathcal P}\int\frac{d^3k}{(2\pi)^3}
   \int_{-\infty}^{\infty}\frac{dk_0}{2\pi}\int_{-\infty}^{\infty}
   \frac{dk_0'}{2\pi}\left[1+f(k_0)+f(k_0')\right]\nonumber\\
   &&
   \frac{2\pi\ \varepsilon(k_0')\ 
   \delta[{k_0'}^2-({\mathbf{k}}-{\mathbf{p}})^2
   -m_\pi^2]\ 2\ {\mbox I}{\mbox m}I^t(k_0,k)}{p_0-k_0-k_0'}\, ,
   \label{reS}
\ee
where ${\mathcal P}$ represents the principal part of the integral. 

The integration in Eq.~(\ref{reS}) can only be performed numerically. 
Figure~3 shows a plot of the temperature-dependent terms of the function 
$\tilde{S}(\tilde{p}_0,p)= -\ (24\pi^4){\mbox R}{\mbox e}S^r(\tilde{p}_0,p)$ 
for $T=m_\pi$ and $\tilde{p}_0$ taken as the solution of 
Eq.~(\ref{dispersion}) up to the $P$-independent terms of Eq.~(\ref{self}). 
>From this figure we see that $\tilde{S}$ is non-vanishing at $p=0$ and thus it 
also contributes to the perturbative increase of the pion mass. We also notice 
that $\tilde{S}$ is a monotonically increasing function of $p$ in the
kinematical range considered. However, this increase is not strong enough to 
cause a qualitatively significant change on the shape of the dispersion 
curve, given that this function contributes multiplied by the perturbative 
parameter $\xi^2$.
%%%%%%%%%%%%%%%%%%%%%%%%%%%%%%%%%%%%%%%%%%%%%%%%%
\vspace{1cm}
\begin{figure}[h!] % fig 3
\hspace{-1cm}
\centerline{\epsfig{file=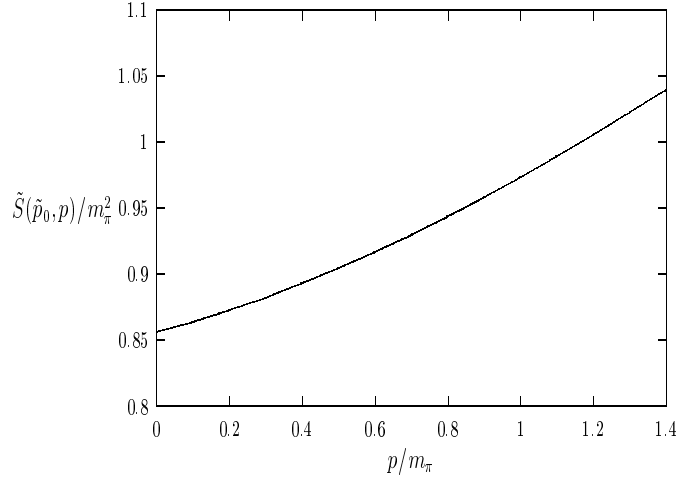,height=2.5in,width=3.5in}
}
\vspace{1cm}
\caption{
$\tilde{S}(\tilde{p}_0,p)/m_\pi^2$ as a function of
		 $p/m_\pi$ for $T=m_\pi$. The function shows a mild
		 monotonical increase in the kinematical region considered.
		 This increase is not enough to change the qualitative
		 behavior of the dispersion curve, given that $\tilde{S}$
		 contributes at second order in the expansion parameter $\xi$.
}
\end{figure}
%%%%%%%%%%%%%%%%%%%%%%%%%%%%%%%%%%%%%%%%%%%%%%%%%

Including all the terms, the dispersion relation up to next to leading order,
for $T\sim m_\pi$ and in the small momentum region is obtained as the 
solution to
\be
   p_0^2=p^2 +
   \left\{1 + \xi g(T/m_\pi) + 
   \frac{\xi^2}{2}g(T/m_\pi)\left[9h(T/m_\pi)-4g(T/m_\pi)\right]
   \right\}m_\pi^2 + \xi^2\tilde{S}(p_0,p)\, .
   \label{disp}
\ee 
%%%%%%%%%%%%%%%%%%%%%%%%%%%%%%%%%%%%%%%%%%%%%%%%%%%%
\vspace{1cm}
\begin{figure}[h!] % fig 4
\hspace{-1cm}
\centerline{\epsfig{file=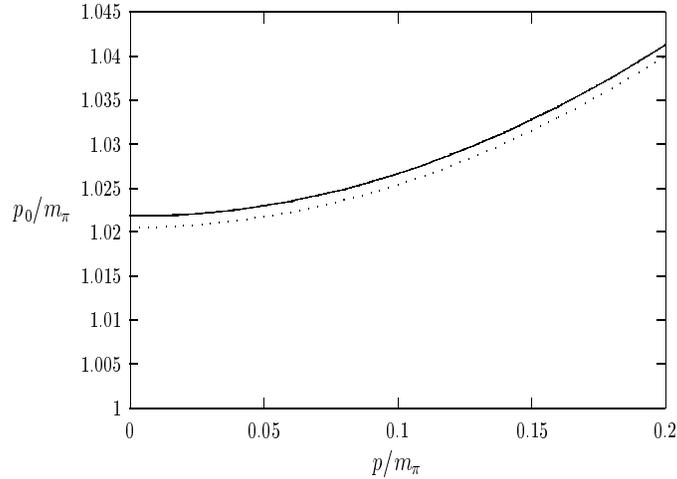,height=2.5in,width=3.5in}
}
\vspace{1cm}
\caption{
Pion dispersion relation obtained as the solution to
		 Eq.~(\ref{disp}) for $T=m_\pi$ (upper curve). Shown is also
		 the dispersion relation obtained by ignoring the term
		 $\tilde{S}(p_0,p)$ (lower curve).
}
\end{figure}
%%%%%%%%%%%%%%%%%%%%%%%%%%%%%%%%%%%%%%%%%%%%%%%%%%%%
Figure~4 shows the dispersion relation obtained from Eq.~(\ref{disp})
for $T=m_\pi$ where we also display the solution without the term 
$\xi^2\tilde{S}(p_0,p)$. As mentioned before, inclusion of this last term does 
not alter the shape of the dispersion relation in this kinematical regime.

In conclusion, we have shown that in the linear sigma model at finite
temperature, the one-loop modification of the sigma propagator induces a 
modification in the one-sigma two-pion and four-pion vertices in such a way as 
to preserve the $\chi$WIs. Strictly speaking, chiral Ward identities fully
constrain vertices only in the limit where all the external momenta are 
zero~\cite{Lee}. Nevertheless, we have also checked that these functions 
arise from considering all of the possible contributions to the vertices of 
interest, when maintaining only the zeroth order terms in an expansion in the 
parameter $(m_\pi/m_\sigma)^2$~\cite{progress}. We have used these objects to 
compute the next to leading order correction to the pion propagator in a pion 
medium for small momentum and for $T\sim m_\pi$. We have shown that the linear 
sigma model yields the same result as $\chi$PT at leading order in the 
parameter $\xi=m_\pi^2/4\pi^2f_\pi^2$ when use is made of a systematic 
expansion in the parameter $(m_\pi/m_\sigma)^2$ at zeroth order. This result 
was to be expected since the kinematical regime we consider is that where the 
temperature, the pion momentum and the pion mass are treated as small 
quantities, such as in the case of $\chi$PT. The main modification to the pion 
dispersion curve in the considered kinematical regime is a perturbative 
increase of the in-medium pion mass. The shape of the curve is not 
significantly altered. For a sigma meson with a mass on the order of 600 MeV 
and in the mentioned kinematical regime, corrections in powers of the 
parameter $(m_\pi/m_\sigma)^2$ are small. If a lighter sigma meson exists, 
corrections of order $(m_\pi/m_\sigma)^2$ could be important. It remains to 
include possible effects introduced by a high nucleonic density. Finally,
it is interesting to note that the formalism thus developed could be employed 
to explore the behavior of the pion dispersion curve in the large momentum 
region where, in principle, the lowest order $\chi$PT Lagrangian cannot be 
used. These issues will be treated in a following up work.

Support for this work has been received in part by CONACyT M\'exico under 
grant numbers I27604-E, 29273-E and 32279-E.

\end{document}